\documentclass[9pt,twocolumn,twoside]{opticajnl}
\journal{opticajournal} 

\usepackage{xcolor}
\definecolor{color2}{RGB}{0,0,0} 

\usepackage{fancyhdr}
\setboolean{shortarticle}{false}

\usepackage{soul}
\usepackage{braket}
\usepackage{amsmath}
\usepackage{lineno}

\newcommand{\murel}{\ensuremath \mu_{\mathrm{rel}}}
\usepackage[colorlinks=true]{hyperref}
\makeatletter
\newcommand\setcurrentname[1]{\def\@currentlabelname{#1}}
\makeatother
\newcommand{\ev}[1]{\ensuremath{\left\langle #1 \right\rangle}}

\DeclareMathOperator{\var}{Var}

\title{Certifying non-classicality and non-Gaussianity through optical parametric amplification}

\author[1,2,*]{Mahmoud Kalash}
\author[1,2]{M.H.M. Passos}
\author[3]{Éva Rácz}
\author[3]{László Ruppert}
\author[3]{Radim Filip}
\author[1,2,$\dagger$]{Maria~V.~Chekhova}

\affil[1]{Max Planck Institute for the Science of Light, Staudtstra\ss{}e 2, 91058 Erlangen, Germany}
\affil[2]{Friedrich-Alexander Universit\"at Erlangen-N\"urnberg, Staudtstra\ss{}e 7/B2, 91058 Erlangen, Germany}
\affil[3]{Department of Optics, Palack\'y University, 17. listopadu 1192/12, 779 00 Olomouc, Czech Republic
}

\affil[*]{mahmoud.kalash@mpl.mpg.de}
\affil[$\dagger$]{maria.chekhova@mpl.mpg.de}
\usepackage{todonotes}
 \newcounter{mycomment}

 \newcounter{mycommentnew}

\newcommand{\remove}[1]{}


\makeatletter
\renewcommand{\titlefont}{\normalfont\rmfamily\bfseries\fontsize{18}{20}\selectfont}
\makeatother

\begin{abstract}
Non-Gaussian states of light are essential for numerous quantum information protocols; thus, certifying non-Gaussianity is crucial. Full quantum state tomography, commonly used for this purpose, 
is a complicated procedure and yields inconclusive results for strongly mixed  states.
Certifying non-Gaussianity through directly measurable parameters is a simpler alternative, typically achieved by 
measuring photon-number probabilities - either directly, using photon-number resolving detectors, or through Hanbury Brown--Twiss type measurements with single-photon detectors.
Here, we demonstrate theoretically and experimentally that optical parametric amplification combined with conventional intensity detectors can effectively replace this approach without the need for photon-number resolution. In our method, we measure the mean photon number and the second-order correlation function for the amplified state. Using it, we successfully certify the non-Gaussianity of a heralded quasi-single-photon state.  Since optical parametric amplification is a broadband and multimode process, our method provides a foundation for developing high-dimensional quantum technologies utilizing broadband multimode non-Gaussian states.
\end{abstract}

\setboolean{displaycopyright}{false} 

\begin{document}

\maketitle

\section{Introduction}\label{Sec:Intro}
Non-classical states of light, i.e., states that cannot be represented as mixtures of coherent states, are indispensable for quantum photonics. 
But several quantum information protocols, such as, for instance, entanglement distillation or quantum error correction, require even more `exotic' non-classical resources, namely quantum non-Gaussian states~\cite{Walschaers2021,Kawasaki2024Nov,Lvovsky2020Jun} -- those which cannot be represented even as mixtures of states with Gaussian Wigner functions.

While there is no experimentally testable definition of nonclassicality (NC) or non-Gaussianity (NG), there are several sufficient conditions (witnesses) for them. Nonclassicality, for instance, can be experimentally certified through anti-bunching ~\cite{Perina2017}. Certifying a state being non-Gaussian is less trivial. 
This may be done through the measurement of the Wigner function: its negativity~\cite{Kenfack2004Aug,Walschaers2017Oct} is a direct evidence of non-Gaussianity. 
However, for mixed non-Gaussian states, which are commonly encountered in practice, the Wigner function can be positive~\cite{Radim2011_NG, Jezek2011Nov,  Jezek2012}. 

To certify the quantum NG or NC of states whose properties are not apparent through the Wigner function, other measurements are needed. One possibility is measuring the value of the Wigner function at the origin by a photon-parity detector, together with the mean photon number~\cite{Genoni2013}. Another, broader used witness, is based on the photon-number (Fock) probabilities of a state~\cite{Radim2011_NG,Jezek2011Nov,Fiurasek_2021,Fiurasek2025,Filip2013_PhysRevA.88.043827}. 
\begin{figure*}[t!]
    \centering
\includegraphics[width=.9\textwidth]{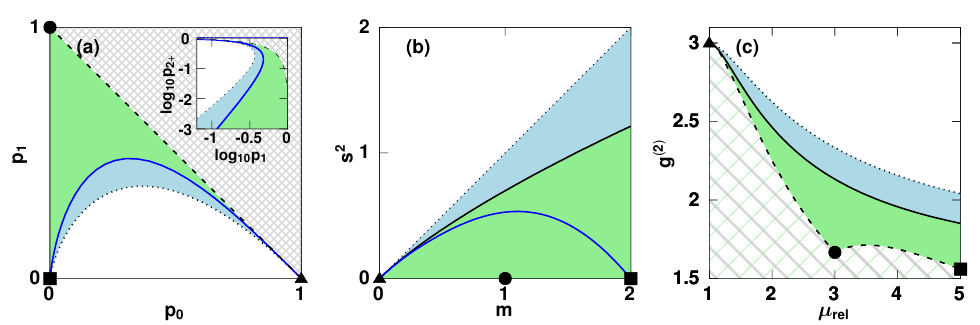}
    \caption{(a) Non-Gaussianity (NG) and non-classicality (NC) witnesses in terms of vacuum (\(p_0\)) and single-photon ($p_1$) probabilities~\cite{Radim2011_NG}. The area above the dotted line (blue and green parts) is not accessible by classical states, while the green area above the blue line is not accessible by Gaussian states. The cross-hatched area is non-physical. Inset: the same criteria in terms of single-photon ($p_1$) and higher photon-number (\(p_{2+}\)) probabilities. (b) The recently proposed~\cite{Racz2025b} NG witness in terms of photon-number mean \(m\) and variance \(s^2\)  (black line). The blue line shows the photon-number probability-based NG witness converted to \(m,s^2\) variables. The dotted line is the boundary of the anti-bunching non-classicality witness \(g^{(2)} < 1\). (c) NG and NC witnesses in terms of the post-amplification second-order correlation function $g^{(2)}$ and mean photon number relative to the one of the amplified vacuum $\mu_{\mathrm{rel}}$, applicable to phase-independent states. The dashed area is where phase-independent states cannot reach. The dashed line for \(\mu_{\mathrm{rel}}\in (1, 3)\) corresponds to different input mixtures of a vacuum and a single photon and for \(\mu_{\mathrm{rel}}\in (3, 5)\), to input mixtures of single-photon and two-photon states. White areas in all panels correspond to points that classical states can reach. The black triangular, circular, and square markers correspond to Fock states $|0\rangle, |1\rangle,$ and $|2\rangle$, respectively.}
    \label{fig:NG-original}
\end{figure*}
Figure~\ref{fig:NG-original}(a) shows how NC and NG can be certified through the probabilities of zero-photon (\(p_0\)) and single-photon (\(p_1\)) events: for classical states and a fixed probability \(p_0\), the probability \(p_1\) cannot exceed a certain limit, which is a function of \(p_0\) (the dotted line). A similar limit (blue solid line) cannot be exceeded for Gaussian states~\cite{Radim2011_NG, Jezek2011Nov, Filip2013_PhysRevA.88.043827}. If \(p_1\) exceeds one of these limits, we can certify the NC or NG of the state.
In particular, the green and blue areas in Fig.~\ref{fig:NG-original}(a) cannot be reached by classical states, and the green area cannot be accessed through any mixture of Gaussian states either (for details, see Sec.~\ref{sec:methods}, \nameref{sec:methods-witness}). The grey hatched areas in Fig.~\ref{fig:NG-original}(a) correspond to non-physical states (\(p_0+p_1>1\)).  Experimentally, \(p_0\) and \(p_1\) can be measured through the standard Hanbury Brown--Twiss (HBT) setup \cite{Jezek2011Nov,Lachman2013Dec} with single-photon detectors, as long as the mean photon number is much less than 1. Applying a similar method to brighter states requires detectors with photon-number resolution or multiple single-photon detectors.

Recently, some of us have shown that a NG witness can be formulated in a form applicable to brighter states~\cite{Racz2025b}, involving the mean number of photons and its variance (Fig.~\ref{fig:NG-original}(b)), similar to the sub-Poissonian witness of NC. The NG is certified whenever the variance is below the black line, which is a stricter condition than sub-Poissonian behavior (below the dashed line), but much less strict than the earlier NG condition (blue line). Still, this witness requires intensity, or photon-number, measurements in the mesoscopic range, which is technically difficult.

To overcome this problem, we propose to amplify the input state using a phase-sensitive optical parametric amplifier (OPA), also known as a noiseless amplifier~\cite{Levenson1993}. Phase-sensitive amplification increases the efficiency of direct detection~\cite{Bencheikh1995,MarandiOPA2024}; it also dramatically improves quantum state reconstruction~\cite{Leonhardt1994} and enables quantum state characterization without homodyne detection, through direct detection only~\cite{Shaked2018Feb,Frascella2019Sep,Nehra2022Sep,Kalash2023,Racz2024}, including the simultaneous measurement of squeezing in multiple eigenmodes~\cite{Barakat2025}. 

Here, we use an OPA with direct intensity detection to certify the NC and NG of quantum states. Specifically, we show that the post-amplification measurement of mean photon number and second-order correlation function using conventional intensity detectors without photon-number resolution can effectively replace coincidence measurements and deal with  
 both faint and bright states.

To prove the principle, we seed an OPA with a heralded quasi-single-photon state and study the photon-number statistics.
Although the full photon-number distribution of the amplified state generally may be used for state tomography \cite{Kalash2023}, in our case, it does not suffice to reveal NG. Due to the significant degradation of the input single photon and multiphoton effects stemming from the conditional preparation of the state, no Wigner negativity is present, a situation frequently observed under realistic experimental conditions.
We show, however, that our witness based on the mean photon number and second-order correlation function is robust enough to certify NG even for highly degraded states. This is demonstrated by tracking the witness as we increase the brightness of the input state, revealing a gradual transition from certified NG to certified NC only, and eventually to inconclusive result as higher-order photon contributions arise. The results align well with those obtained from the standard coincidence measurements for faint heralded single-photon states 
and extend to brighter states, where single-photon detectors become unusable.

\section{Results}\label{Sec:Theory}

\begin{figure*}[!t]
\centering
\includegraphics[width=1\linewidth]{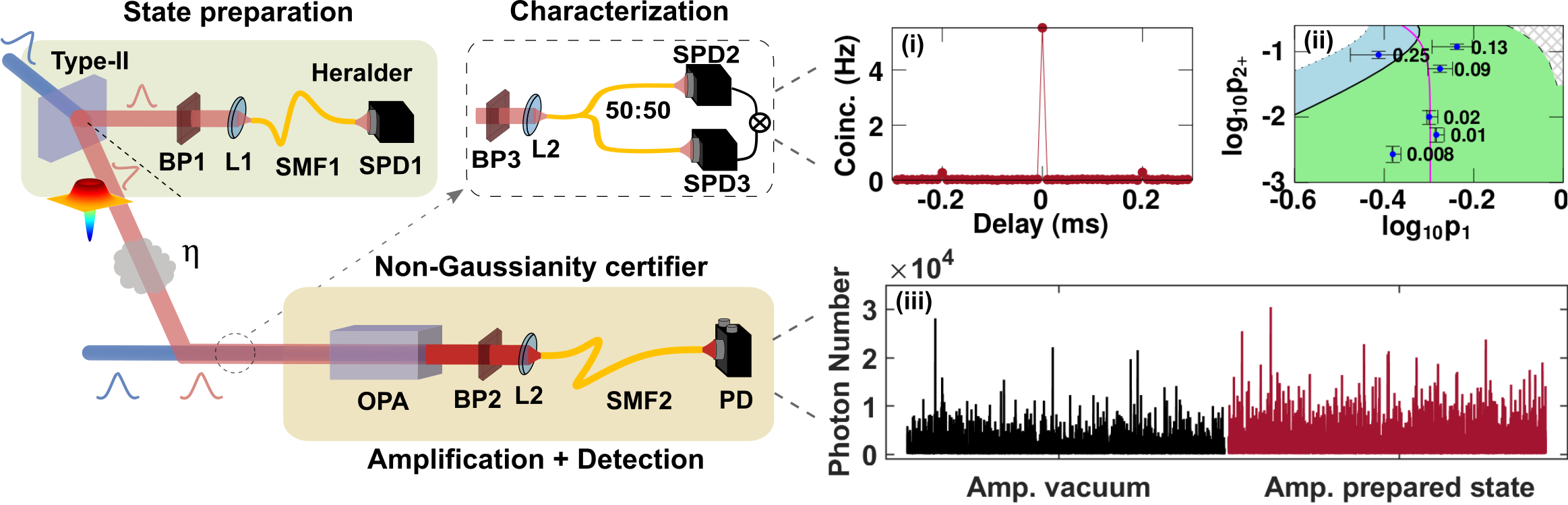}
     \caption{ Experimental scheme. Quantum states are prepared (green shaded box) through heralding on the idler photons of type-II SPDC using a single-photon detector (SPD1) after filtering with a band-pass filter (BP1) and a single-mode fibre (SMF1). For certification (yellow shaded box), heralded signal photons are fed into  optical parametric amplifier OPA, whose output, after similar filtering with BP2 and SMF2, is registered by photodetector PD (a camera). The efficiency $\eta$ includes both the detection and coupling into the OPA. The dashed box shows an HBT setup with single-photon detectors SPD2,3 used for the characterization of the heralded state. Inset (i): the histogram of coincidences between SPD1 and SPD2/SPD3 for the brightness of the source, measured by SPD1, set at 0.02 photons per pulse at the source output. Inset (ii): single- and multiphoton probabilities obtained for heralded states of varying brightness, quantified through the mean number of photons per pulse at the source output. The magenta line shows the analytic curve for the fixed heralding efficiency of 51\%. Inset (iii): photon numbers for a sequence of {$35000$} pulses in the cases of vacuum (left) and heralded photon (right) states at the input.}
     \label{setup-main}
\end{figure*}

\subsection*{Photon-number statistics of amplified quantum states}

When characterizing a state at the input of an OPA, one parameter to consider is the mean photon number after amplification, $\mu\equiv\left< \hat{N}\right>$, with $\hat{N}=\hat{a}^{\dagger}\hat{a}$ being the photon number operator and $\hat{a}^{\dagger}$ ($\hat{a}$)  the photon creation (annihilation) operator, respectively. Using the Bogoliubov transformation, 
\begin{equation}
    \label{eq:boboliubov}
    \hat{a} =\hat{a}_0\cosh{G} + \hat{a}_0^{\dagger}\sinh{G}, 
\end{equation}
with the parametric gain
$G$ and input operators $\hat{a}_0,\hat{a}_0^\dagger$, we can evaluate $\mu$ for various input states. For example, when the OPA is seeded with a vacuum, a thermal state with the mean photon number $\overline{n}$, and a Fock state $|n\rangle$, the post-amplification mean photon numbers are, respectively, $\mu_0=\sinh^2{G}$, $\mu_{\mathrm{th}} = (2 \overline{n} + 1) \sinh^2{G}$, and  $\mu_{|n\rangle} = (2n+1)\sinh^2{G} + n \approx (2n+1)\sinh^2{G}$~\cite{Kim1989Sep}.
Although sensitive to the input energy, the latter two cases demonstrate that the mean photon number after amplification is, on its own, not suitable to discriminate between different types of states: $\mu$ is the same for thermal or Fock inputs when $\overline{n}=n$. 

On the other hand, the normalized second-order correlation function {$g^{(2)}(0)=\frac{\braket{a^{\dagger}a^{\dagger}aa}}{{\braket{a^{\dagger}a}\braket{a^{\dagger}a}}}$, here denoted for simplicity $g^{(2)}$, 
has proven to be a highly effective tool for characterizing the intensity statistics of an electric field, thereby providing insights into the nature of photon streams and their statistical properties~\cite{ Mandel1995Sep}.
For instance, $g^{(2)}=1$ corresponds to coherent (Poissonian) statistics, while $g^{(2)}=2$ indicates thermal (super-Poissonian) statistics. To illustrate how $g^{(2)}$ reflects the input state NG, consider amplification of different states.
At high amplification gain, vacuum and thermal states both yield $g^{(2)}\approx3$, due to strong quadrature anti-squeezing causing large intensity fluctuations \cite{Kim1989Sep}.
However, amplified Fock states yield significantly lower $g^{(2)}$  ($\leq1.66 \ \text{for} \ n \geq 1$), although they have the same quadrature anti-squeezing. Since anti-squeezing conserves the nature of the initial quadrature distribution, this drop in $g^{(2)}$ clearly indicates sensitivity to the NG of the input states.

\subsection*{Non-classicality and non-Gaussianity witnesses through optical parametric amplification}

In the current method, we measure the intensity and its moments after the amplification of a state. 
 Instead of the anti-bunching NC witness (\(g^{(2)} < 1\), Fig.~\ref{fig:NG-original}(b)), we construct similar witnesses in terms of the mean photon number $\mu$ and the normalized second-order correlation $g^{(2)}$ after amplification (Fig.~\ref{fig:NG-original}(c)). 
To make the result independent of the gain \(G\), 
we define the normalized mean photon number 
\(\mu_{\mathrm{rel}} \equiv \mu / \mu_{\mathrm{vac}}\), 
where \(\mu_{\mathrm{vac}} = \sinh^{2} G\) is the mean photon number 
for amplified vacuum.
For sufficiently high gain, \(G>3\), these new witnesses do not depend on \(G\) and converge to the asymptotic witnesses, which are valid in general. Meanwhile, for weaker amplification (\(G<3\)), we can obtain less demanding witnesses (that is, a higher value of \(g^{(2)}\) is sufficient for the same relative mean, see Suppl.~\ref{sec:NG-proof}).

We plotted the boundaries of these new witnesses (solid and dotted black lines in Fig.~\ref{fig:NG-original}(c)) as parametric curves, see Sec.~\ref{sec:methods}, \nameref{sec:methods-witness}.
Note that the use of these witnesses is only possible for states whose Wigner function is phase-independent because we amplify a single quadrature and, as a result, get information about this specific quadrature only. 
Witnesses can also be constructed for the general case, but are outside of the scope of the current manuscript.
 
The grey dashed area in Fig.~\ref{fig:NG-original}(c) cannot be reached by phase-independent states. That is, a measurement result in that area indicates the phase-dependence of the input state. For example, a weakly squeezed single photon state, which is phase-dependent and non-Gaussian, can reach point ($\murel=2, \,g^{(2)}=1.66$) within the grey dashed area in Fig.~\ref{fig:NG-original}(c). The dashed line bounding this area corresponds to input mixtures of vacuum and single-photon states for \(\mu_{\mathrm{rel}} \in [1, 3)\) and to input mixtures of single- and two-photon states for \(\mu_{\mathrm{rel}} \in [3, 5)\), which are, of course, non-Gaussian. 

\remove{On the other hand, amplification maps the first and second photon-number moments $m=\sum_n np_n$ and $s^2= \sum_n n^2 p_n$ that hold information about these probabilities, to the output moments $\mu$ and $\sigma^2$, as well as the $g^{(2)}$ as [\textcolor{red}{methods}]
\begin{equation}
    \begin{split}
        \murel=\mu&/\mu_v=(2m+1),\\
        g^{(2)}=\frac{\sigma^2}{\mu^2}-\frac{1}{\mu}+1&,
        \;\sigma^2 =3\sigma_v^2 s^2 + \frac{m^2+m+1}{2}, 
    \end{split}\label{outputmoments}
\end{equation}
with $\mu_v$ and $\sigma^2_v$ are the moments corresponding to amplified vacuum, and $\murel$ is introduced for convenience. Equations (\ref{outputmoments}) show that Gaussian statistics, which impose constraints on $p_1$, similarly restrict $\mu$ and $g^{(2)}$ for amplified states, thereby allowing the certification of non-Gaussianity through \textcolor{blue}{a direct measurement after amplification}.}

\begin{figure*}[t!]
\centering
\includegraphics{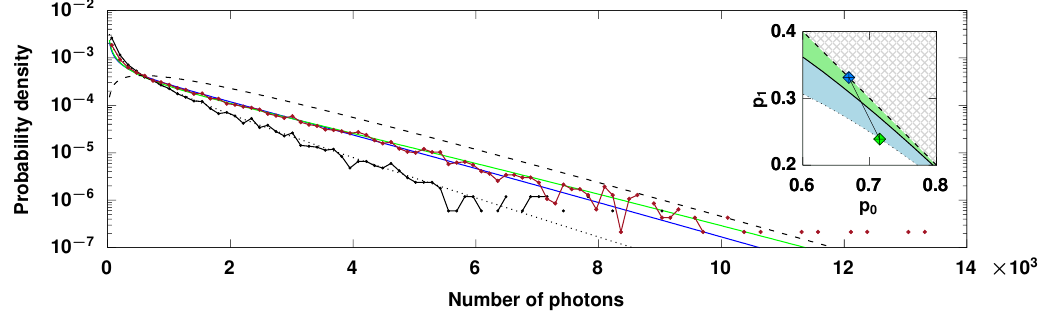}
         \caption{Photon-number distributions of the amplified vacuum (black solid line) and amplified heralded state with a brightness of 0.1 photon per pulse ({red} 
         solid line), measured over 35000 pulses. The dotted/dashed black line shows the theoretical photon-number distribution of amplified vacuum/ideally amplified single photon. The blue and green lines show the theoretical distributions for input states with $p_0=0.67,p_1=0.33, p_{2+}=0$ (non-Gaussian) and $p_0=0.715,p_1=0.239,p_{2+}=0.046$ (non-classical), respectively. Inset: points corresponding to these states (blue and green crosses, respectively) in the ($p_0, p_1$) diagram.} 
    \label{example}
\end{figure*}

\subsection*{Special case: a heralded quasi-single-photon state}

For demonstration, we focus on the case of a quantum state obtained through single-photon heralding at the output of a weakly pumped two-mode squeezer. This state is a mixture of a vacuum, a single-photon state, and higher-order Fock states whose contributions are small ($p_{2+}\ll 1$). 
This heralded state is straightforward to model (see Suppl.~\ref{sec:heralded-model}), and all photon-number probabilities and moments are analytically accessible for a given set of input parameters (losses, mean number of photon pairs leaving the crystal). This scenario is therefore ideal to compare the performance of the coincidence- and moment-based witnesses.

\paragraph{Experiment.}
We generate the heralded quasi-single-photon state from a twin-beam source and track its NG through optical parametric amplification while varying the source brightness. Higher brightness leads to increased multi-photon contributions, which in turn reduces the NG of the state. Figure \ref{setup-main} shows the experimental scheme. The source is pumped with 1 ps pulses at 400 nm and phase-matched for non-collinear degenerate 
type-II SPDC. The brightness of the source, i.e., the number of photons per pulse (ppp), is controlled by adjusting the input pump power.  The idler beam is directed to a heralding single-photon detector (SPD1), while the signal is fed into the OPA. The latter is a nonlinear crystal cut for type-I collinear degenerate phase matching and pumped with the same laser pulses but at higher power. To ensure single-mode detection, we use a 0.5-nm spectral filter (BP1,2) and a single-mode fiber (SMF1,2) in both heralding and detection channels. Finally, the  photodetection (PD) is done with a camera. More details about the experimental setup, the source, and the OPA are presented in Sec.~\ref{sec:methods}, \nameref{sec:experimental-scheme}.}  

\paragraph{Characterization of the source.}
We start by characterizing the source through the standard HBT coincidence measurement \cite{Jezek2011Nov}. For this, we employ a three-fold coincidence scheme: a single-photon detector in the heralding channel (SPD1) triggers a coincidence circuit formed by two detectors (SPD2 and SPD3) in the signal arm. To measure the heralding efficiency, the coincidences between SPD1 and SPD2/SPD3 at $0\,\mu\mathrm s$ delay are detected (Fig.~\ref{setup-main}(i)). The brightness of the source is set at 0.02 ppp, with the latter measured by SPD1 and evaluated at the source output. Side peaks at \(\pm 200\, \mu\mathrm s\) are caused by neighboring pulses and represent accidental coincidences.
 For this measurement, we use a 10-nm bandpass filter BP3 in the heralded arm, which ensures the detection of signal photons across all spectral modes. After correcting for the system optical transmissivity and SPDs' efficiencies, the heralding efficiency amounts to $\eta_{H}=51\pm2\%$ (see  Sec.~\ref{sec:methods}, \nameref{sec:heralding-efficiency}). This defines the purity of the heralded state at the input of the amplifier. 
 
 The rates of triggered coincidences between SPD2 and SPD3 provide the probabilities $p_0$, $p_1$, and $p_{2+}$ (see Sec.~\ref{sec:methods}, \nameref{sec:Fock-from-HBT}), from which the NG witness of the heralded state can be evaluated. Fig.~\ref{setup-main}(ii) shows the obtained $p_1$ and $p_{2+}$ values while varying the source brightness between 0.008 and 0.25 ppp. We see that the source features NG for brightness levels up to 0.13 ppp, but loses it as the brightness increases. The magenta line shows the theoretical values (see Suppl.~\ref{sec:heralded-model}) of \(p_1\) and \(p_{2+}\) under a variable source brightness and a fixed $\eta_H= 51\%$, and aligns reasonably well with the experimental data.

\begin{figure*}[t!]
\centering
\includegraphics{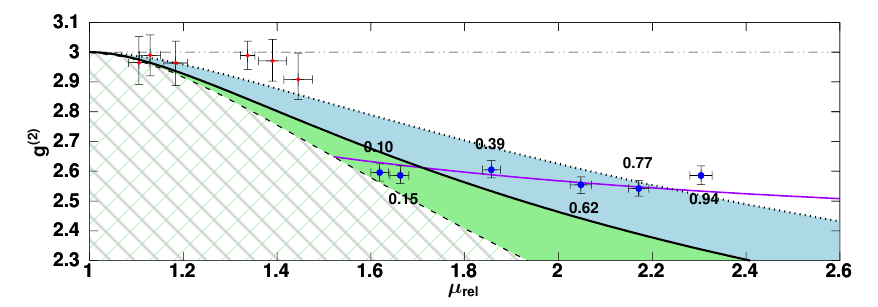}
    \caption{\label{fig:g2ng} Experimental post-amplification \(g^{(2)}\) and relative mean photon numbers $\mu_{\mathrm{rel}}$. Blue circles correspond to heralded states for different source brightnesses, quantified through mean number of photons per pulse at the output. The blue and green areas correspond to the NC and NG witnesses from Fig.~\ref{fig:NG-original}(c). 
        The magenta  line shows the dependence calculated analytically for a fixed effective transmittance of \(26\%\) while increasing the brightness of the source (from left to right).
    Red points show the results for a thermal state obtained from the same source without heralding. The measurement shows \(g^{(2)}\) close to the theoretical value 3 ({black dash-dotted} line) and no nonclassicality. 
       }
\end{figure*} 
\begin{figure}[!b]
    \centering
    \includegraphics{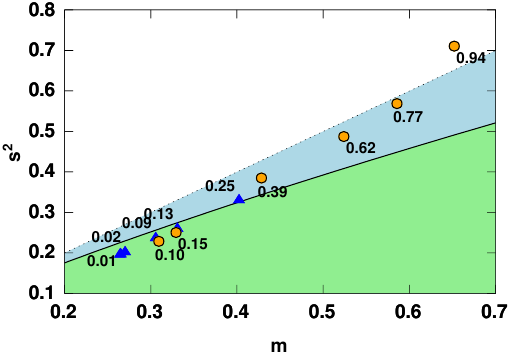}
    \caption{Estimated photon-number mean and variance values from the coincidence measurement (blue triangles) and the amplified measurement (orange disks). The numeric values next to the symbols correspond to the mean number of photons per pulse at the source output. The solid black line shows the boundary of the non-Gaussianity witness, the dotted black line is the boundary of the non-classicality (anti-bunching) witness, \(s^2 = m\).}
    \label{fig:cmp}
\end{figure}
\paragraph{Parametric amplification of the examined states.}
We now feed the examined states into the OPA and record photon numbers for 35000 pulses conditioned on detection events at SPD1. Figure~\ref{setup-main}(iii) shows an example of these records when the amplifier is seeded by vacuum (left) and by a heralded single-photon state with the brightness set at 0.1 ppp (right). In the latter case, the mean number of photons is clearly increased. Processing the data also shows that $g^{(2)}$ is reduced: while for amplified vacuum it is measured to be $3.00\pm0.03$, for the amplified heralded state it becomes $2.58\pm0.03$. This reduced value is, however, higher than the 1.66 expected for an amplified pure single-photon state \cite{Kim1989Sep}. Similarly, the mean photon number also increases insufficiently, resulting in $\murel=1.66\pm0.02$ instead of the expected value of 3. These discrepancies are caused by the reduced purity 
\(\eta = \eta_H \eta_{\mathrm{OPA}}\) of the examined state, 
where \(\eta_{\mathrm{OPA}}\) quantifies the efficiency of amplification 
and is primarily degraded by imperfect spectral/spatial mode matching between the input and amplifier modes (see Sec.~\ref{sec:methods}, \nameref{sec:mode-matching}). 

The corresponding intensity histograms 
$\mathcal P(N)$  are shown in Fig.~\ref{example} (for the derivation of the continuous approximations of the number distribution of squeezed Fock states, see Suppl.~\ref{sec:cont-approx}). The black dashed line presents the theoretical continuous approximation of the number distribution $\mathcal P_1(N)$ of an amplified pure single photon ($p_0=p_{2+}=0$, $p_1=1$), while the black dotted line shows the theoretical distribution $\mathcal P_0(N)$ of an amplified vacuum  ($p_1=p_{2+}=0$, $p_0=1$) and it perfectly fits the corresponding experimental distribution (black solid line). The measured intensity histogram of the amplified heralded state is shown by a red solid line. As a mixture, this distribution can be approximated by $\mathcal P(N)=p_0\mathcal P_0(N)+p_1\mathcal P_1(N)+p_{2+}\mathcal P_{2}(N)$ (with $\mathcal P_2(N)$ denoting the approximate intensity distribution of a squeezed Fock-2 state), and it lies in between $\mathcal P_0(N)$ and $\mathcal P_1(N)$ for negligible $p_{2+}$ values. To illustrate the impact of higher Fock contributions, we compare two distributions that yield the same experimental mean photon number. The blue line represents a pure vacuum–single-photon mixture ($p_0=0.67$, $p_1=0.33$, $p_{2+}=0$), which is a non-Gaussian state by definition. The corresponding blue cross in  $p_0, p_1$ space (inset in Fig.~\ref{example}) sits precisely at the boundary between non-Gaussian and non-physical states. In contrast, the green solid line depicts a slightly more complex mixture including few higher Fock contributions ($p_0=0.715,p_1=0.239,p_{2+}=0.046$). This state lies exactly on the NC boundary (green cross in the inset in Fig.~\ref{example}). Despite describing very different states, the blue and green distributions both fit the experimental data reasonably well. This is mainly because the purity of the amplified state is degraded. This comparison highlights a key point: for low-purity states,  non-Gaussianity is non-trivial to certify through the post-amplification photon-number distribution. 

On the other hand, the use of $\murel$ and $g^{(2)}$ (Fig.~\ref{fig:g2ng}) confirms the NG of states with low brightness, which are non-Gaussian according to the preliminary state characterization (Fig.\ref{setup-main}(ii)). 
The leftmost data point in Fig.~\ref{fig:g2ng}, corresponding to the 0.1-ppp case, lies well within the NG region (green area). 
As the brightness of the input state increases (0.15, 0.39, 0.62, 0.77, 0.94 ppp), the contribution from higher-order Fock states becomes more significant, gradually pushing the measured $(\murel, g^{(2)})$ values out of the NG and then out of the NC region. The results are consistent with an overall transmissivity of $\eta=26\%$ (magenta line). Given a heralding efficiency of 51\%, this implies an amplification efficiency of $\eta_{\mathrm{OPA}}=51\%$.

We also examined a set of thermal states (red points), obtained by measurements without conditioning on the events from SPD1. These states are Gaussian and classical. From left to right, the points correspond to the same source brightness levels as used in the conditional measurements. As expected, the points lie in the region where neither NG nor NC can be confirmed. The measured $(\murel, g^{(2)})$ values are close to $g^{(2)}=3$ (black dash-dotted line).

We can now compare these new results with the ones of coincidence measurement. To this end, we convert both sets of data -- probabilities $p_1,\,p_0$ on the one hand and $g^{(2)},\,\murel$ on the other hand - to pre-amplification moments $m,s^2$ (rescaling the coincidence measurement to the average 26\% effective transmittance obtained in Fig.~\ref{fig:g2ng}). The result is shown in Fig.~\ref{fig:cmp}. We see that the two sets of points line up very well, and the points with similar mean numbers of photons lie close to each other. The measurement with amplification was mostly done on brighter states to demonstrate the broader applicability of the scheme.

\section{Discussion}
We have proposed and experimentally 
demonstrated optical parametric amplification as a tool for certifying the non-Gaussianity and non-classicality of quantum states. Specifically, we theoretically showed that the directly measurable mean photon number and second-order correlation function after amplification provide information similar to that revealed from photon-number probabilities and coincidence measurements. 

Experimentally, we examined a heralded quasi-single-photon state generated through spontaneous parametric down-conversion and successfully tracked its transition from the non-Gaussian to the non-classical and then to classical regime by increasing its brightness. We found that although the photon-number distributions of amplified Gaussian and non-Gaussian states are barely distinguishable, our proposed method allows us to accurately characterize these states. Furthermore, we showed that losses, including those arising due to mismatch between the modes of the examined state and the amplifier, are not critical for this measurement and can always be corrected for. Our experimental results closely match the predictions of the developed theory and agree with the results obtained through the standard coincidence measurements.

In addition to eliminating the need for single-photon detectors, an OPA naturally enables multimode detection. This is especially important for applications where distinct non-Gaussian states occupying multiple modes need to be certified simultaneously. Thus, our method is an excellent candidate for high-dimensional quantum information technologies.

This OPA-based approach to certifying non-classicality and non-Gaussianity can also be generalized to phase-dependent states; however, this requires 
a more involved setup \cite{Racz2025b}.

We note that a current, independent approach came to our attention, which also uses correlation functions to characterize the non-Gaussianity of states \cite{Hotter2025}. Similarly to the standard approach, they also use single-photon detectors, while our method only needs intensity detectors.

\section{Methods}\label{sec:methods}

\subsection*{Detailed experimental scheme}
\begin{figure}[t]
\centering
\includegraphics[width=1\linewidth]{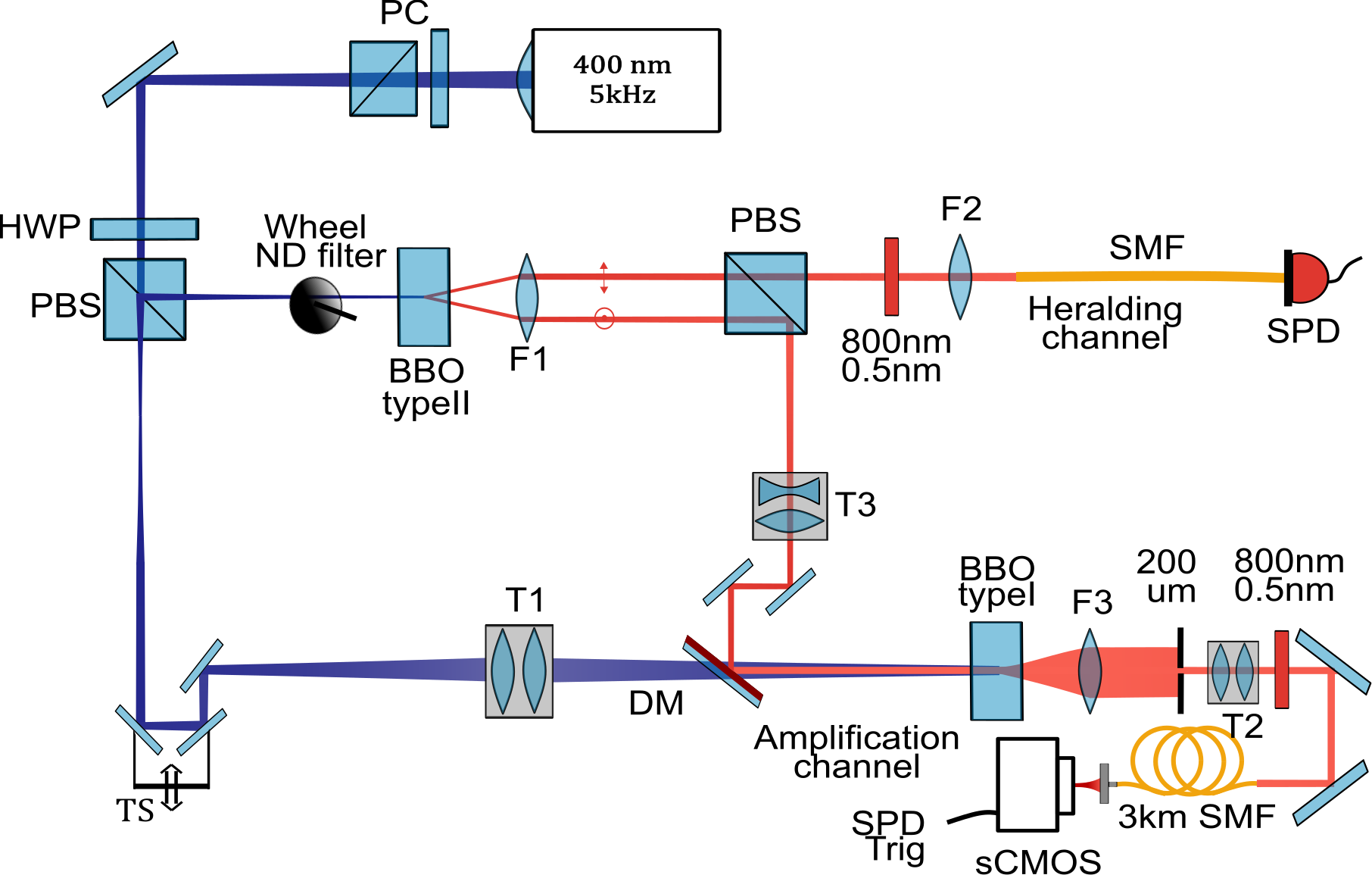}
     \caption{Full experimental setup. HWP, half-wave plates; M1-4, mirrors; DM, dichroic mirror; PBS, polarizing beam splitter; ND, Neutral Density Filter; F1-3, Lenses; T1-3, Telescopes; TS, Translational Stage; SM, spherical mirror; BBO, Beta-Barium Borate; SMF, Single Mode Fiber; sCMOS, scientific Complementary Metal–Oxide–Semiconductor camera.}
    \label{setup}
\end{figure}

\setcurrentname{Detailed experimental scheme}\label{sec:experimental-scheme} Figure~\ref{setup} presents the complete experimental setup. After power control (PC) with a polarizing beam-splitter and a half-wave plate, another polarizing beam splitter (PBS) separates the pump beam (1 ps pulses at $400$ nm at a 5-kHz repetition rate) 
into two arms. The pump power splitting ratio is controlled with a half-wave plate (HWP). In the upper arm, the pump is focused to waist $w_p=105\,\mu\mathrm m$ onto a 3-mm thick beta barium borate (BBO) crystal, configured for beam-like~~\cite{Kurtsiefer2001Nov,Takeuchi2001Jun} type-II SPDC, generating  photon pairs at the degenerate wavelength of 800 nm. A variable neutral density (ND) filter is included to vary the brightness of the SPDC source by adjusting the pump power. The idler photons (H-polarized) are transmitted through the PBS and directed to the heralding arm. After collimation with lens F1 = 30\,$\mathrm{mm}$, the idler beam passes through a 0.5-nm interference filter centered at 800 nm to ensure spectral purity and is then coupled into a single-mode fiber (SMF) using lens F2 $ = 5.5\,\mathrm{mm}$. The SMF guides the photons to a superconducting nanowire single-photon detector (SPD) with a quantum efficiency of approximately 80\%. This detector acts as a heralding trigger, sending a signal to open the shutter of an sCMOS camera for synchronized detection of the amplified signal. The signal photons (V-polarized), reflected by the PBS, are directed into the amplification arm.

 The same laser source used for SPDC also pumps the OPA, which is a 3-mm-thick bismuth borate (BiBO) crystal  under collinear degenerate type-I phase matching. The amplifier is operated in a phase-sensitive configuration with the parametric gain $G = 6.5\pm0.5$, optimized for high signal-to-noise ratio. To achieve spatial mode matching, telescopes T1 (-50 mm  and 50 mm) and T3 (-30 mm and 75 mm) are employed. Temporal overlap between the signal and pump pulses is ensured using a translation stage (TS) in the pump path. The amplified signal is collimated by lens F3=75 mm and filtered: spatially, with a 200-$\mu$m pinhole, and spectrally, with a 0.5-nm interference filter similar to the one in the heralding arm. The amplified radiation is detected by an sCMOS camera whose quantum efficiency is approximately 50$\%$. Upon receiving a trigger signal from the SPD, the camera requires $\sim\!10\ \mu s$ to start the exposure. To synchronize it with the heralding signal, the amplified beam is delayed by $15\ \mu s$ using a 3-km SMF.
\subsection*{Mode matching and the efficiency of amplification}\setcurrentname{Mode matching and the efficiency of amplification}\label{sec:mode-matching}
Mode matching is an important factor governing the efficiency of the OPA process. Indeed, for a single-mode field $a_0$ at the input of the OPA, the addressed state is $a=\sqrt{\eta_{\mathrm{OPA}}} a_0  + \sqrt{1-\eta_{\mathrm{OPA}}} a_v$, where $a_v$ is the vacuum field and $\eta_{\mathrm{OPA}}$ defines the mode overlap between the input and the amplifier mode. That is, if mode matching is imperfect ($\eta_{\mathrm{OPA}}<1$), the efficiency of the amplification decreases because of the vacuum contribution. In our case, the heralded photon, generated using a type-II source, has a much narrower spectral mode when compared to the amplifier mode based on type-I collinear-degenerate phase matching. This is the main reason behind the degraded OPA efficiency, which amounts to $\eta_{\mathrm{OPA}}=51\%$. However, this issue can be resolved by spectrally shaping the OPA modes. Alternatively, a broadband single-photon source based on type-I parametric down-conversion would suit the detection scheme.
\subsection*{Heralding efficiency}\setcurrentname{Heralding efficiency}\label{sec:heralding-efficiency} 

The heralding efficiency $\eta_H$ is defined as the probability to observe a photon in the signal arm, provided that a photon was detected in the heralding arm. If higher Fock-state probabilities are negligible, this quantity coincides with the single-photon probability \(p_1\).
For instance, setting the brightness of the source to 0.01 ppp, we had $6\cdot 10^{-4}$ coincidences per pulse between SPD1 and SPD2, with $8\cdot 10^{-3}$ clicks per pulse on the heralding arm SPD1. Correcting for 38\% SPD2 efficiency, 50\% beamsplitter, 90\% optical transmissivity, and 85\% fiber coupling, we obtained $\eta_H=51 \pm \ 2$ \%, and it corresponds to the purity of the prepared quasi-single-photon state.
For higher source brightness, multiphoton effects start to play a significant role. Still, one can use the model described in Suppl.~\ref{sec:heralded-model} to evaluate the heralding efficiency, which provides similar numbers.

\subsection*{Photon-number probabilities from HBT measurements}\setcurrentname{Fock probabilities from HBT measurements}\label{sec:Fock-from-HBT}
Assuming that an incoming signal with unknown photon-number distribution is split on a beam splitter with transmittance \(T\), and the transmitted photons are measured by detector \(A\) with probability \(p_A\), whereas the reflected photons are detected by detector \(B\) with probability \(p_B\), we can estimate the photon-number probabilities of a weak signal as
\begin{align}
\label{eq:p2}    p_{2+} &= \frac{Q_2}{2 p_A p_B TR},\\
\label{eq:p1}
p_1 &= 
\frac{Q_1
-p_2
\left[2TR(q_Bp_A+q_Ap_B)+
T^2(1-q_A^2)+R^2(1-q_B^2)\right]
}
{Tp_A+R p_B},\\
\label{eq:p0} p_0 &= 1-p_1 -p_{2+},
\end{align}
with \(Q_2\) denoting the probability of double clicks and \(Q_1\) denoting the probability of single clicks; \(R \equiv 1-T\), \(q_A \equiv 1-p_A\), and \(q_B \equiv 1-p_B\) were introduced for convenience (for details on the derivation of these formulas, see Suppl.~\ref{sec:imperfect-detection}). A similar approach was used in \cite{Jezek2011Nov} to estimate Fock probabilities, the difference is that we also account for imperfect detection (\(p_A \neq 1\) and \(p_B\neq 1\)).

Knowing the value of extra loss characterized by a transmittance \(T'\) 
and suffered before the beam splitter of the HBT scheme, it is straightforward to correct for it assuming a weak signal:
\begin{align*}
    p'_{2+} &= \frac{p_{2+}}{T'^2};\\
    p'_1 &= \frac{p_{1}}{T'}-p'_{2+}\cdot 2(1-T');\\
    p'_0 &= 1 - p'_1 - p'_{2+}.
\end{align*}
These formulas were used to correct the HBT results for optical losses and fiber coupling losses in Fig.~\ref{setup-main}(c). Note that an alternative approach has been recently used in  \cite{Checchinato2024}, where they include the loss directly into the witness.

\subsection*{Non-classicality and non-Gaussianity witnesses}\setcurrentname{Non-classicality and non-Gaussianity witnesses}\label{sec:methods-witness}

The NG witness introduced in \cite{Jezek2011Nov} can be expressed as a parametric curve (solid black line in Fig.~\ref{fig:NG-original}(a)) parametrized by \(r\geqslant 0\):
\begin{equation}
\begin{split}
    \tilde p_0 (r) &=\frac{\exp\{-\frac{1}{4}(e^{4r}-1)(1-\tanh{r})\}}{\cosh{r}}
    ,\\
    \tilde p_1 (r) &=\frac{(e^{4r}-1)\exp\{-\frac{1}{4}(e^{4r}-1)(1-\tanh{r})\}}{4\cosh^3{r}}
.\label{eq:p0p1-criterion}
\end{split}
\end{equation}
Any state for which the pair of zero-photon and single-photon probabilities lies beyond this curve (green area in Fig.~\ref{fig:NG-original}(a)) cannot be represented as a mixture of Gaussian states. The non-classicality witness in terms of number probabilities can be given simply as \(p_1 > -p_0\ln p_0\) \cite{Jezek2011Nov} (area above the dotted line in Fig.~\ref{fig:NG-original}(a)).

For our purposes, we are interested in the post-amplification witnesses for phase-independent input states. It is straightforward to show that for phase-independent states, the asymptotic (large amplification) relationship between pre- and post-amplification moments is quite simple (see Suppl.~\ref{sec:NG-proof}):
\begin{equation}
    \begin{split}
        \murel &= 2m+1, \\
        \sigma^2_{\mathrm{rel}} &= 2\left(1 + m + m^2 + 3s^2\right),
    \end{split}
\end{equation}
where the relative values are normalized to \(\sinh^2 G\) and \(\sinh^4G\), respectively. That is, the value of \(m\) determines \(\murel\), and for \(m\) fixed, \(\sigma^2_{\mathrm{rel}}\) is a linearly increasing function of \(s^2\), and therefore the problem of minimizing \(\sigma^2_{\mathrm{rel}}\) for a fixed value of \(\murel\) is equivalent to minimizing \(s^2\) for a fixed value of \(m\). The problem of minimizing \(s^2\) over arbitrary mixtures of Gaussian states for a fixed value of \(m\) has been solved in \cite{Racz2025b}; the  boundary of this NG witness is shown in Fig.~\ref{fig:NG-original}(b) with a thick black line. Note that for a fixed $m$, the minimization of the variance $s^2$ is equivalent to minimizing the pre-amplification values of the second-order correlation function, the second moment of the integrated intensity, and the non-centered second moment of the photon numbers, so such alternatives can be used interchangeably. (Note that the latter is a convex function, and that form is used to prove the validity of witness for mixtures, for details, see \cite{Racz2025b}.) 

Furthermore, because for large amplification, \(g^{(2)} = 1+\sigma^2_{\mathrm{rel}}/\murel^2\), the boundary of the post-amplification non-Gaussianity witness can be given as a parametric curve (see Fig.~\ref{fig:NG-original}(c), thick black line) 
\begin{equation}
    \begin{split}
    \label{eq:NG-amplified}
        \tilde \mu_{\mathrm{rel}} (r) &= e^r\cosh 2r,\\
        \tilde g^{(2)}(r) &= 3 - 3 \cdot\frac{\sinh ^2r \cdot \left(\sinh 2 r+1\right)}{\cosh^2 2 r},
    \end{split}
\end{equation}
with \(r \geqslant 0\). For large values of \(r\), \(\tilde g^{(2)}(r) \approx 3/2 + 9e^{-4r}\), 
whereas for small values, \(\tilde g^{(2)}(r) \approx 3 - 3r^2\).
The non-Gaussianity witness for a given value of \(\mu_{\mathrm{rel}}\) is then
\begin{equation}
g^{(2)} < \tilde g^{(2)}\left(\left[\tilde \mu_{\mathrm{rel}}\right]^{-1}\!(\mu_\mathrm{rel})\right),
\end{equation}
with \([\cdot]^{-1}\) denoting the inverse of a function. The inverse exists since \(\tilde \mu_{\mathrm{rel}}(r)\) is a strictly increasing function. The asymptotic boundary is reached very quickly in the value of the amplification parameter \(G\), and the boundary of the witness for non-asymptotic values lies higher than the asymptotic curve (see Suppl.~\ref{sec:NG-proof}). As a consequence, using the asymptotic curve as a witness is a conservative choice.

The non-classicality witness in terms of \(g^{(2)}\) and \(\mu_{\mathrm{rel}}\) is straightforward to calculate using that the photon number variance of a coherent state is equal to its mean (\(s^2 = m\)), and yields
\begin{equation}
    \label{eq:NCnew}
    g^{(2)} < \frac32 +\frac{3}{\murel} -\frac {3}{2\murel^2}.
\end{equation}

In their stated forms, the NC/NG witnesses are valid only for a single mode. However, this assumption is close to the truth in our case, since the parametrically amplified scheme includes a single-mode fiber.

\begin{backmatter}
\bmsection{Funding} 
The project/research is part of the Munich Quantum Valley, which is supported by
the Bavarian state government with funds from the Hightech Agenda Bavaria.  M. V. C. acknowledges
funding from the Deutsche Forschungsgemeinschaft (grant number CH
1591/16-1). This research was funded within the QuantERA II Programme (project SPARQL), which has received funding from the European Union’s 
Horizon 2020 research and innovation programme under Grant Agreement No 101017733. We also acknowledge the grant 23-06224S of the Czech Science Foundation, European Union’s HORIZON Research and Innovation Actions under Grant Agreement no. 101080173 (CLUSTEC) and the Quantera project CLUSSTAR (8C24003) of MEYS, Czech Republic.

\bmsection{Acknowledgments} R.F.\ acknowledges discussions with Ch. Hotter and A.S. Sørensen, who independently derived quantum non-Gaussian criteria for correlation functions.

\bmsection{Disclosures} The authors declare no conflicts of interest.

\bmsection{Supplemental document}

\end{backmatter}

\bibliography{References}

\bibliographyfullrefs{References}

\newpage
\onecolumn
\appendix
\begin{center}
{\Huge\bfseries Supplementary document}
\end{center}
\renewcommand\thefigure{S.\Roman{figure}}    
\setcounter{figure}{0}
\renewcommand\thetable{S.\Roman{table}}    
\setcounter{table}{0}
\renewcommand\theequation{S.\arabic{equation}}    
\setcounter{equation}{0}

\section{Post-amplification non-Gaussianity and non-classicality witnesses}\label{sec:NG-proof}

{
In this section, we show how to derive the post-amplification non-classicality and non-Gaussianity witnesses for phase-independent states, that is, states whose Wigner function only depends on \(x^2+p^2\). The general expressions for the first two pre-amplification photon number moments are
\begin{align*}
    m &\equiv \ev{\hat N} = \int W(x, p)\left(x^2+p^2\right) \, \mathrm dx\, \mathrm dp -\frac 1 2,\\
    s^2 + m^2 &= \ev{\hat N^2} = \int W(x, p)\left(x^2+p^2\right)\left(x^2+p^2 -1 \right) \, \mathrm dx\, \mathrm dp,
\end{align*}
with \(s^2\) denoting the photon number variance. By switching to polar coordinates in the second integral and using the phase-independence assumption, it is straightforward to show that
\begin{align}
    \label{eq:mpre}  m &= 2\ev{\hat X^2} - \frac 1 2 , \\
    \label{eq:s2pre} s^2 &= \frac 8 3 \var \hat X^2 -\frac 43 \ev{\hat X^2}^2 - \frac 1 4.
\end{align}
At the same time, assuming that the parametric gain is large enough to neglect the contribution of the squeezed quadrature to the photon numbers (\(\hat N_{\mathrm{post}}\approx e^{2G}\hat X^2\)), we have for the post-amplification moments
\begin{align}
    \label{eq:mupost}  \mu &\approx e^{2G}\ev{\hat X^2} \approx 4\mu_0 \ev{\hat X^2}, \\
    \label{eq:sigma2post} \sigma^2 &\approx e^{4G}\var \hat X^2 \approx 16\mu_0^2 \var \hat X^2,
\end{align}
with \(\mu_0 = \sinh^2 G \approx e^{2G}/4\) denoting the mean photon number of squeezed vacuum.

Rearranging Eqs.~(\ref{eq:mpre}-\ref{eq:sigma2post}), we obtain
\begin{align}
\label{eq:murel0}\mu_{\mathrm{rel}}&\equiv \frac \mu {\mu_0} \approx 4\ev{\hat X^2} =   1 + 2m,\\
\label{eq:sigma2rel}\sigma^2_{\mathrm{rel}} &\equiv \frac {\sigma^2} {\mu_0^2}  \approx 16\var{\hat X^2} = 2\left(1  + m + m^2 + 3s^2\right),
\end{align}

These equations show that the post-amplification relative mean is a linear function of the pre-amplification relative mean. That is, fixing the value of \(\murel\) also fixes the value of \(m\). Furthermore, for a fixed value of \(m\), \(\sigma^2_{\mathrm{rel}}\) is also a linear function of \(s^2\). Therefore the problem of finding the minimum of \(\sigma^2_{\mathrm{rel}}\) with \(\murel\) fixed is equivalent to finding the minimum of \(s^2\) with \(m\) fixed among all mixtures of Gaussian states. The latter problem has been solved in \cite{Racz2025b}.
Consequently, we can also express such witnesses through \(\mu_{\mathrm{rel}}\) and \(g^{(2)}_{\mathrm{post}} \approx 1 + \sigma^2_{\mathrm{rel}}/\mu_{\mathrm{rel}}^2\) by substituting the functions \(m_{\mathrm{NG}}(r)\) and \(s^2_{\mathrm{NG}}(r)\) from \cite{Racz2025b} into \eqref{eq:murel0} and \eqref{eq:sigma2rel}.

Figure \ref{fig:gain-dependence} shows how the border of the non-Gaussianity witness behaves for different values of the parametric gain \(G\) (we used the same calculations as detailed above, except using the exact post-amplification moments of the phase-randomized versions of the optimal displaced squeezed vacua derived in \cite{Racz2025b} instead of the approximations in \eqref{eq:mupost}-\eqref{eq:sigma2post}): \(\mu(G, r) = \left(\left(e^{-2 r}+e^{6 r}\right) \cosh (2 G)-2\right)/4\), \(\sigma^2(G, r) = e^{-4 (G+r)} \left(\left(e^{4 G}-1\right)^2 e^{16 r}+4 \left(2 e^{4 G}+5 e^{8 G}+5\right) e^{8 r}-8 e^{4 (G+r)} (3 \cosh (4 G)+5)+2 e^{4 G}+7 e^{8 G}+7\right)/128\)). It shows that the border essentially coincides with the asymptotic curve for any \(G \geqslant 3\). Furthermore, the border for smaller values of \(G\) lies higher than the asymptotic curve; that is, the witness for asymptotic values of \(G\) is stricter than the witness for small values of \(G\). Consequently, if a measurement result lies below the asymptotic curve, it is certainly meets the non-Gaussianity criterion for all smaller values of the parametric gain.

\begin{figure}[hp]
    \centering
    \includegraphics[width=0.5\textwidth]{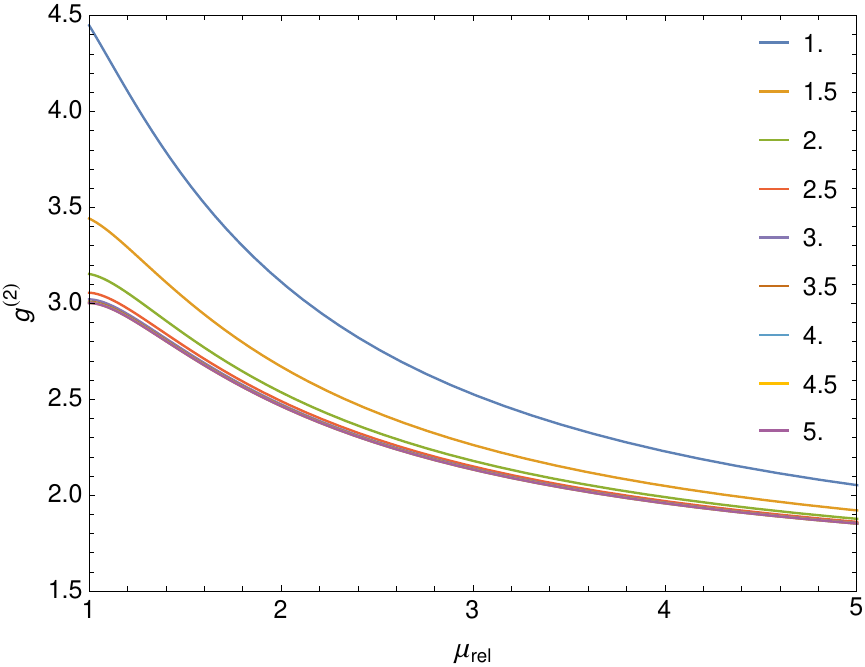}
    \caption{Border of the non-Gaussianity witness for different values of the parametric gain \(G\) (\(G\in [1,5]\), see plot legend).\label{fig:gain-dependence}}
\end{figure}

The dashed line in Fig.~\ref{fig:NG-original}(c) represents the border of points that can be reached by phase-independent states after squeezing. For deriving the analytic form of this border, we use that all phase-independent states can be expressed as mixtures of Fock states, and therefore  \(s^2 = \sum_n n^2 p_n - m^2\) with \(m = \sum n\cdot p_n\). Using the convexity of \(n^2\) and Jensen's inequality, it follows
 that for a given mean \(m \in [n, n+1) \Leftrightarrow \mu_{\mathrm{rel}}\in[2n+1, 2n+3) \), the minimal variance \(s^2\) (and therefore \(\sigma^2_{\mathrm{rel}}\) and \(g^{(2)}\)) is reached by a mixture of Fock-\(n\) and Fock-\((n+1)\). 
That is, the equation of the dashed line is
\begin{align*}
\mu_{\mathrm{rel}} \in [1, 3):&\quad \mu_{\mathrm{rel}}(p) = \mu_0^{-1}\left[p\mu_0 + (1-p)\mu_1)\right];\quad
g^{(2)}(p) = \mu_0^{-2}\left[p(\mu_0^2+\sigma_0^2) + (1-p)(\mu_1^2+\sigma_1^2)\right]; \quad 0 \leqslant p < 1 \\
\mu_{\mathrm{rel}} \in [3, 5):&\quad \mu_{\mathrm{rel}}(p) = \mu_0^{-1}\left[p\mu_1 + (1-p)\mu_2)\right];\quad
g^{(2)}(p) = \mu_0^{-2}\left[p(\mu_1^2+\sigma_1^2) + (1-p)(\mu_2^2+\sigma_2^2)\right]; \quad 0 \leqslant p < 1,
\end{align*}
with  \(\mu_n \equiv \sinh^2 G + n\cosh 2G\) and \(\sigma^2_n \equiv \sinh^2(2G)\cdot(n^2+n+1)/2\) denoting the mean and the variance of the squeezed Fock-\(n\) state \cite{Kim1989Jul}.
}

\section{Model of the heralded state}\label{sec:heralded-model}
In this section, we describe the calculations for a thermal model to approximate the different statistics in our heralding scheme.
Assuming the source that emits thermally distributed pairs of photons with mean \(m_0 = q_0/(1-q_0)\), we show how to derive all the relevant quantities of the heralded state. The most useful mathematical tool here is the probability generating function (PGF), defined for an arbitrary non-negative integer valued random variable \(X\) as \(G_X(z) \equiv \langle z^X \rangle = \sum_{n = 0}^{\infty} \mathbb P (X = n)\cdot z^n\), with \(\mathbb P(\cdot)\) denoting the probability of an event. The PGF of the thermal distribution is 
\[G_0(z) \equiv \langle z^{N_0}\rangle = \sum_{n = 0}^{\infty} (1-q_0)q_0^n\cdot z^n = \frac{1-q_0}{1-q_0z},\]
with \(N_0\) denoting the number of pairs of photons leaving the source.

Let us first calculate the PGF corresponding to the number of pairs of photons at the source provided the heralding detector had a click in the general case, \(G_0(z) = \sum_n p_n z^n\):
\begin{align*}
G_{N_0|H>0}(z) &= \sum_{n = 0}^\infty z^n \cdot \mathbb P\left(N_0 = n \mid  H > 0\right) = \sum_{n = 0}^\infty z^n \cdot \frac{\mathbb P\left(N_0 = n \text{ and }  H > 0\right)}{\mathbb P(H > 0)} \\
&= 
\sum_{n = 0}^\infty z^n \cdot \frac{\mathbb P\left(N_0 = n\right) - \mathbb P\left(N_0 = n\text{ and }  H = 0\right)}{1 - \mathbb P(H = 0)}\\
&= \sum_{n = 0}^\infty z^n \cdot \frac{p_n - p_n(1-\eta_H)^n}{1 - \sum_{k = 0}^{\infty}p_k(1-\eta_H)^k} = \frac{G_0(z) - G_0\left[(1-\eta_H)z\right]}{1 - G_0(1-\eta_H)},
\end{align*}
where \(H\) is zero the heralding detector does not click and positive otherwise (note that this model of the detector cannot distinguish between a single photon or more photons), \(\eta_H\) denotes the aggregate transmissivity at the heralding arm including detector efficiency.
Plugging in the specific form of \(G_0\) corresponding to thermal distribution into this formula, we get 
\[  G_{N_0|H>0}\left(z\right) =
z \cdot
\frac{1-q_0}{1-q_0 z}\cdot 
\frac{1-q_0 (1-\eta _H) }
{1-q_0 (1 - \eta _H) z}.
\]
To obtain the heralded signal \(S\), we need to introduce further losses, characterized by an aggregate transmissivity \(\eta_S\). An ideal beam splitter with transmissivity \(\eta\) transforms an original PGF \(G\) as \(\tilde G(z) = \sum_{n = 0}^{\infty} p_n \sum_{k = 0}^n \binom{n}{k} \eta^k(1-\eta)^{n-k} \cdot z^k =  \sum_{n = 0}^{\infty} p_n (1-\eta + \eta z)^n = G(1-\eta + \eta z)\). Therefore the PGF of the heralded state can be given as
\[
    G_{S|H>0}(z) = G_{N_0|H>0}\left(1-\eta_S + \eta_S\cdot z\right) = 
    \left(1-\eta_S + \eta_S\cdot z\right) \cdot
\frac{1-q_0}{1-q_0 \left(1-\eta_S + \eta_S\cdot z\right)}
\cdot 
\frac{1-q_0 (1-\eta _H) }
{1-q_0 (1 - \eta _H) \left(1-\eta_S + \eta_S\cdot z\right)}.\]
For comparison, the generating function of an ideal heralded single photon (where all multiphoton effects are neglected) is simply \(1-\eta_S + \eta_S\cdot z\).

A PGF \(G(z) = \sum_{n=0} p_n z^n\) in general provides a full description of the distribution: it allows to access to individual probabilities (\(p_k = {G^{(k)}(0)}/{k!}\)) and moments (\(m_k = \frac{\mathrm d^k}{\mathrm d s^k} G(e^s)|_{s = 0}\)). Specifically, \(m = G'(1)\) and \(s^2 = G''(1) + G'(1) - \left[G'(1)\right]^2\). 

Finally, to obtain the heralding efficiency $\eta_S$, we fit the parameters of this model ($q_0$, $\eta_H$ and $\eta_S$) to match the measured coincidence probabilities (detailed in Suppl.\ \ref{sec:imperfect-detection}).

\section{Continuous approximation of the number distribution of squeezed Fock states}\label{sec:cont-approx}

In this section, we show a simple continuous approximation of the number distribution of squeezed Fock states applicable in the case of strong squeezing, where the contribution of the squeezed quadrature is neglibible. Therefore, we can simply use the distribution of the square of the anti-squeezed quadrature as a proxy for the photon number.

\begin{figure}[!t]
\centering
\includegraphics[width=.7\textwidth]{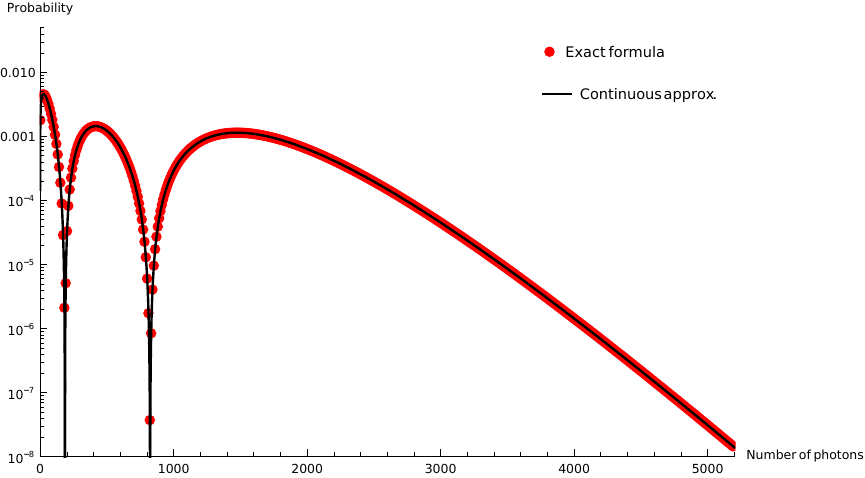}
\caption{Comparison of the exact formula from \cite{Kim1989Sep} and the continuous approximation derived above for \(n = 5\) and \(G = 3\). Note that in order to get an approximation of the discrete number probabilities, the previously derived formula has to be multiplied by 2 to account for the fact that every second number probability is zero according to the exact formula.\label{fig:cont-approx}}
\end{figure}

By expanding the Laguerre polynomial, we obtain the following form of the Wigner function of the Fock state \(\left|n\right\rangle\):
\begin{equation*}
W_n(x,p) = \frac{(-1)^n}{\pi} e^{-x^2 - p^2}\mathcal L_n\left[2(x^2 + p^2)\right]
= \frac{(-1)^n}{\pi} e^{-x^2 - p^2}\sum_{i = 0}^n
\binom{n}{i}\frac{(-2)^i}{i!}\sum_{j = 0}^i\binom{i}{j}x^{2(i-j)}p^{2j}.
\end{equation*}
From this, the Wigner function of the squeezed Fock state is 
\begin{equation*}
W_n(G; x,p) = W_n\left(e^{-G}x, e^Gp\right)
= \frac{(-1)^n}{\pi} e^{-e^{-2G}x^2 - e^{2G}p^2}\sum_{i = 0}^n
\binom{n}{i}\frac{(-2)^i}{i!}\sum_{j = 0}^i\binom{i}{j}(e^{-G}x)^{2(i-j)}(e^Gp)^{2j}.
\end{equation*}
The quadrature PDF is then simply
\begin{equation*}
f_n(G; x) = \int_{-\infty}^{\infty} W_n\left(G; x, p\right)\, \mathrm d p 
= \frac{e^{-G}}{\sqrt \pi}e^{-G} e^{-e^{-2G}x^2}
\sum_{i = 0}^n \binom{n}{i}\frac{(-2)^i}{i!}
\sum_{j = 0}^i\binom{i}{j}\left(e^{-G}x\right)^{2(i-j)}\frac{(2j)!}{4^j j!},
\end{equation*}
where we used the formula for the even moments of the standard normal distribution (\(M_{2j} = (2j-1)!! \equiv \frac{(2j)!}{j!2^j}\)).
From this, by the change of variables \(N \equiv x^2/2\), we can obtain the continuous approximation of the photon number distribution as
\begin{equation}
\label{eq:cont-approx}
\mathcal P_n(G; N) = \sqrt{\frac 2 N} f_n\left(G; \sqrt{2N}\right)
=\frac{1}{\sqrt{\theta \pi}}N^{-1/2} e^{-N/\theta} \times (-1)^n
\sum_{i = 0}^n \binom{n}{i}\frac{(-2)^i}{i!}
\sum_{j = 0}^i\binom{i}{j}\frac{(2j)!}{4^j j!}\left(\frac N \theta \right)^{i-j}
\end{equation}
with \(\theta \equiv e^{2G}/2\). This can be factorized as $\mathcal P_n(G; N)= \mathcal P_0(G; N) \times R_n(N/\theta)$. 
That is, we have the intensity PDF of squeezed vacuum (gamma distribution) 
\[\mathcal P_0(G; N) = \frac{1}{\sqrt{\theta \pi}}N^{-1/2} e^{-N/\theta}\]
multiplied by a degree-\(n\) polynomial of \(N/\theta\),
\[R_n(t) = \sum_{i = 0}^n\sum_{j = 0}^i \binom{n}{i}\frac{(-2)^i}{i!}
\binom{i}{j}\frac{(2j)!}{4^j j!}t^{i-j}.\]
The first few of these polynomials $R$ are shown in Table \ref{tab:Rn}.
For example,  $\mathcal P_2(G; N)= \mathcal P_0(G; N) \times R_2(N/\theta) = \frac{1}{\sqrt{\theta \pi}}N^{-1/2} e^{-N/\theta} \times (1 - 2N/\theta)^2/2$. As an example we  compare the exact, discrete formula for the photon number distribution of a squeezed Fock-5 state and its continuous approximation for \(G = 5\) (see Figure \ref{fig:cont-approx}), showing that the given approximation is very close to its discrete counterpart.

\begin{table}[hptb]
\centering
\caption{Polynomials \(R_n(t)\) from the continuous approximation of the number distribution of squeezed Fock states. \label{tab:Rn}}
\renewcommand{\arraystretch}{1.5}
\begin{tabular}{c|c|c|c|c|c|c}
    \(n\) & 0 & 1 & 2 & 3 & 4 & 5\\
     \hline
    \(R_n(t)\) & 1 & \(2t\) &  \((1-2 t)^2/2\)  & \(2t(3-2 t)^2/3! \) & \((3 - 12 t + 4 t^2)^2/4!\) & \(2t(15 - 20 t + 4 t^2)^2/5!\)
\end{tabular}
\end{table}

\section{Correction for imperfect detection in the coincidence measurement}\label{sec:imperfect-detection}

In this section, we will show how we accounted for imperfect detection during the evaluation of the coincidence measurement. We assume that an incoming signal with unknown number distribution is split on a beam splitter with transmittance \(T\). The transmitted photons are measured by detector \(A\) with probability \(p_A\), whereas the reflected photons are detected by detector \(B\) with probability \(p_B\). Assuming that the state is quite dim, \(p_{3+} = 0\), we can estimate its number probabilities using the measured proportion of double detection \(Q_2\) and single detection \(Q_1\).

The probability of not registering any photon on detector \(A\) is
\begin{equation*}
\mathbb P \left(A = 0\right) = \sum _{n = 0}^{\infty}p_n\sum_{m = 0}^n \binom{n}{m} T^m(1-T)^{n-m}(1-p_A)^m = \sum _{n = 0}^{\infty}p_n(1-Tp_A)^n = M(z_A)
\end{equation*}
with \(M(z)\equiv \sum_{n = 0}^{\infty} p_n z^n\) denoting the probability generating function of the number of photons in the signal, and \(z_A \equiv 1-Tp_A\). It is similarly simple to show that \(\mathbb P(B = 0) = M(z_B)\) and \(\mathbb P \left(A = 0 \wedge B = 0\right) = M(z_{AB})\), with \(z_B \equiv 1 - (1-T)p_B\), and \(z_{AB} \equiv 1 - Tp_A -(1-T)p_B\). Therefore,
\begin{align}
\label{eq:Q2}    Q_2 &= \mathbb P\left(A > 0 \wedge B > 0\right) 
= 1 - M(z_A) - M(z_B) + M(z_{AB}) \\
\label{eq:Q1}    Q_1 &= \mathbb P\left(A > 0 \wedge B = 0 \vee A = 0 \wedge B > 0\right) = M(z_A) + M(z_B)- 2 M(z_{AB}) 
\end{align}
Supposing that the signal is a mixture of Fock states 0 to 2 (at least approximately), we have \(M(z) = p_0 + p_1 + p_2z^2\) with the constraint \(p_0 + p_1 + p_2 = 1\). Substituting this form into Eqs.~\ref{eq:Q2}-\ref{eq:Q1}, and also using the constraint on the sum of probabilities, we have a straightforward set of three equations for the three unknown quantities \(p_0\), \(p_1\), and \(p_2\); its solution is shown in Eqs.~(2)-(4) in the main text.

\end{document}